\documentclass[conference]{IEEEtran}
\IEEEoverridecommandlockouts
\usepackage{cite}
\usepackage{amsmath,amssymb,amsfonts}
\usepackage{algorithmic}
\usepackage{graphicx}
\usepackage{textcomp}
\usepackage{xcolor}
\usepackage{caption}
\usepackage{subcaption}
\usepackage{multirow}
\usepackage[flushleft]{threeparttable}
\def\BibTeX{{\rm B\kern-.05em{\sc i\kern-.025em b}\kern-.08em
    T\kern-.1667em\lower.7ex\hbox{E}\kern-.125emX}}
\usepackage{array}
\newcolumntype{C}[1]{>{\centering\arraybackslash}m{#1}}
\renewcommand{\baselinestretch}{0.945}

\begin{document}

\title{A Machine Learning based Hybrid Receiver for 5G NR PRACH}

\author{\IEEEauthorblockN{Rohit Singh\IEEEauthorrefmark{1},
Anil Kumar Yerrapragada\IEEEauthorrefmark{2},
Radha Krishna Ganti\IEEEauthorrefmark{3}}
\IEEEauthorblockA{Department of Electrical Engineering\\
Indian Institute of Technology
Madras \\ Chennai, India  600036\\
Email: \IEEEauthorrefmark{1}rohitsingh@smail.iitm.ac.in,
        \IEEEauthorrefmark{2}anilkumar@5gtbiitm.in,
	\IEEEauthorrefmark{3}rganti@ee.iitm.ac.in
}
}

\makeatletter
\def\ps@IEEEtitlepagestyle{%
\def\@oddfoot{\mycopyrightnotice}%
\def\@evenfoot{}%
}
\def\mycopyrightnotice{%
\gdef\mycopyrightnotice{}
}

\maketitle

\begin{abstract}
Random Access is a critical procedure using which a User Equipment (UE) identifies itself to a Base Station (BS). Random Access starts with the UE transmitting a random preamble on the Physical Random Access Channel (PRACH). In a conventional BS receiver, the UE's specific preamble is identified by correlation with all the possible preambles. The PRACH signal is also used to estimate the timing advance which is induced by propagation delay. Correlation-based receivers suffer from false peaks and missed detection in scenarios dominated by high fading and low signal-to-noise ratio. This paper describes the design of a hybrid receiver that consists of an AI/ML model for preamble detection followed by conventional peak detection for the Timing Advance estimation. The proposed receiver combines the Power Delay Profiles of correlation windows across multiple antennas and uses the combination as input to a Neural Network model. The model predicts the presence or absence of a user in a particular preamble window, after which the timing advance is estimated by peak detection. Results show superior performance of the hybrid receiver compared to conventional receivers both for simulated and real hardware-captured datasets. 
\end{abstract}


\section{Introduction}
5G use cases such as Massive Machine Type Communications, have to support a wide range and a large number of UE devices, including mobile phones and various IoT devices. Each of these devices has to attach to a BS, upon turning on. In 5G systems, the BS continuously transmits broadcast signals, which the UE has to detect in order to trigger its initial attachment procedure. The procedure starts with the UE selecting a Random Access Preamble Index (RAPID) which it uses to rotate a known base sequence. Since the BS has limited knowledge about the UE's location and channel conditions, it uses a correlation-based receiver to decode the RAPID and estimate the propagation delay induced by the channel. Delay estimates fed back to the UE can be used to advance future uplink transmissions. Henceforth, we refer to this delay as Timing Advance (TA). 

Initially, the UE transmits on the PRACH with a low power and ramps it by $3/6$ dB in the subsequent transmissions if the previous reception fails. PRACH decoding failures, in the form of false peaks and missed detections are common in correlation-based receivers, particularly in high fading and low Signal-to-Noise Ratio (SNR) scenarios. Note that each retransmission on the PRACH causes a wastage of radio and power resources and increases the latency of attachment. 

To improve the robustness of PRACH decoding (and therefore reduction in PRACH retransmissions), in this paper, we show the design of a hybrid receiver that relies on an AI/ML model and a peak detection module. The input to the model is the Power Delay Profile (PDP) of the correlation, combined across multiple antennas. We have shown that this hybrid receiver performs better than both conventional correlation-based approaches as well as other AI/ML-based approaches. The main contributions of this paper are:
\begin{itemize}
    \item Design of a hybrid receiver for PRACH, which outperforms correlation-based receivers. This work is a lower complexity enhancement of our previous work which relied on large Neural Network (NN) models~\cite{singh2024enhancements}.
    \item Performance analysis of the proposed receiver with Multiple Input Multiple Output (MIMO).
    \item Performance analysis of the proposed receiver on real data.
    \item Explainability insights on the decision making of the model using Shapley Additive Explanations.
\end{itemize}

\section{Background on 5G NR Physical Random Access Channel}
This section provides basic details on the Random Access (RA) procedures and PRACH signal generation for 5G NR. Additionally, it describes the prior literature on both conventional and AI/ML-based receivers. 

\subsection{5G Random Access Procedures}
5G supports two types of RA Procedures - Contention-based RA and Contention-free RA. In contention-based access, the UE is free to choose any random preamble from a set. In Contention-free RA, the UE is pre-assigned the preamble that it has to transmit. This paper focuses on the Contention-based RA. In the first step, the UE chooses a preamble randomly and sends it on the PRACH. This message is known as Msg1. At the BS, the receiver determines which preamble the user has chosen and sends a response, also known as a RA Response (RAR, Msg 2). Uplink resources for the following message from the UE are also allocated using Msg 2. The RAPID is sent back to the UE, where it is compared with the one it had chosen previously. If both match, the UE transmits the next message in the uplink data channel (Msg 3). This is followed by a downlink contention resolution (Msg 4). If the RAPID is correctly detected, the RA procedure succeeds. Otherwise, the UE starts the whole process again, which adds to the latency in the attachment of the UE to the BS. Reducing the initial attach latency is the main motivation for designing a robust receiver for the PRACH. 

\subsection{PRACH Sequence Generation and Transmission}
This paper focuses on the PRACH signaling in Msg1 of the RA process. Considering an OFDM system, the received Msg 1 signal, $\tilde{y}_{u,v,i}(k)$ in the $k^{th}$ resource element, at the $i^{th}$ BS antenna is given by, 

\begin{equation}
    \tilde{y}_{u,v,i}(k) = h_i(k)y_{u, v}(k)e^{\frac{-j2\pi TA_{i}}{N}} + w_{i}(k), 
    \label{eq: RX_PRACH_with_TA}  
\end{equation}

where $h_i(k)$ represents the fading channel experienced by the signal received at the $i^{th}$ antenna, $N$ is the OFDM FFT size, and $w_i(k)$ is the Additive White Gaussian Noise (AWGN) at the $i^{th}$ antenna. $TA_{i}$ is the timing delay experienced by the signal received at the $i^{th}$ antenna. The sequence $y_{u, v}(k)$ is the transmitted frequency domain signal (mapped to the OFDM resource grid using the tables 6.3.3.2-2 to 6.3.3.2-4 given in TS 38.211~\cite{3gpp_38_211}) and it is given by, 
\begin{equation}
    \begin{split}
    y_{u,v}(k) &= \sum_{n=1}^{L_{RA} - 1} x_{u,v}(n) e^{\frac{-j2\pi nk }{L_{RA}}} = X_u(k) e^{\frac{-j2\pi k C_v }{L_{RA}}},
    \end{split}
    \label{eq: prach_dft}
\end{equation}

where the length of the sequence $L_{RA}$, depends on the PRACH preamble format given by Table 6.3.3.1-2 in TS 38.211~\cite{3gpp_38_211}. The initial sequence number $u$ is selected from Table 6.3.3.1-4 in TS 38.211. $x_{u}(n)$ is a Zadoff-Chu base sequence and is given by $x_{u}(n) = e^{\frac{-j\pi un(n+1)}{L_{RA}}}, n = 0,1,.., L_{RA} -1$. Note that $x_{u}(n)$ is first generated in the time domain, upon which a random cyclic rotation is applied to obtain $x_{u,v}(n) = x_u((n + C_v) \text{ mod } L_{RA})$. The cyclic rotation is a function of the preamble index and a parameter $N_{CS}$, which is given in Table 6.3.3.1-7 in TS 38.211. The table indices are configured by higher layers. $X_u(k)$ is the $L_{RA}$-point Discrete Fourier Transform of the base sequence $x_{u}(n)$ and is known at both the transmitter and the receiver. 

In this paper, we use the following commonly configured parameters for illustration: $L_{RA}=139$ and $N_{CS}=13$ resulting in $C_{v} = vN_{CS}$ where $v$ indicates the preamble index ranging from $0,.., \left\lfloor \frac{L_{RA}}{N_{CS}} \right\rfloor-1$, for a base sequence. Note that $L_{RA} = 139$ is more widely used than $L_{RA} = 839$ given its lower frequency domain resource utilization.

The 5G standards define 64 preambles. Note that for the values configured above, $\left\lfloor \frac{L_{RA}}{N_{CS}}\right\rfloor = 10$. Therefore each base sequence can accommodate $10$ RAPID values (corresponding to $10$ preambles), resulting in a total of $7$ base sequences.

Note that Eq.~\eqref{eq: RX_PRACH_with_TA} includes one rotation that uniquely identifies the RAPID (through $C_{v}$ and $u$) and a second rotation that is induced by the propagation delay due to the distance between the UE and the base station. The channel response also adds delay to the signal, due to various multipaths and this delay depends on the location of higher power taps. For an OFDM FFT size $N=4096$, and a sub-carrier spacing of $30$ kHz, a propagation delay of $\frac{4096}{139} = 29.4$ samples (caused by a distance of $71$m) causes the correlation peak at the receiver to shift by one sample.  

\subsection{Existing receivers for 5G NR PRACH}
This section provides a summary of existing work on both conventional and AI/ML-based receiver implementations for PRACH. For ease of understanding the research landscape for PRACH, we provide a summary of key prior works along with a comparison with our proposed receiver in Table~\ref{tab:comparison_table}.

\subsubsection{Conventional approaches}
Conventional receivers use correlation~\cite{pham2019proposed, kamata2021detection} followed by a thresholding mechanism~\cite{hu2012method, threshold_2},~\cite{multi_threshold_journal} to detect the user. Typical implementations achieve correlation by an equivalent IDFT operation. In the case of multiple receiver antennas, correlation values are combined from all the receiver antennas using Equal Gain Combining since the individual channels are not known. The combined correlation values are used either in the complex form (Complex Delay profile, CDP) or in the power form (Power Delay Profile, PDP), as shown below.

\begin{equation}
CDP \triangleq \chi(k) = \frac{1}{N_{RX}}\sum_{i=1}^{N_{RX}} \text{IDFT} \left[  \frac{(\tilde{y}_{u,v,i}(k))}{X_u(k)}\right]
\label{eq: cdp}
\end{equation}

\begin{equation}
PDP \triangleq |\chi(k)|^2 = \frac{1}{N_{RX}} \sum_{i=1}^{N_{RX}} \bigg |\text{IDFT} \left[  \frac{(\tilde{y}_{u,v,i}(k))}{X_u(k)}\right]\bigg |^2
\label{eq: pdp}
\end{equation}

Here, $N_{RX}$ is the number of receiver antennas, $\tilde{y}_{u,v,i}(k)$ is the received frequency domain data at the $i^{th}$ antenna and $k = 0, .., L_{RA} - 1$. Since, the value of $N_{CS}$ is $13$, groups of $13$ values of $k$ correspond to a single preamble window (single RAPID). 

\begin{figure}[h!]
    \centering
    \includegraphics[width=0.49\textwidth]{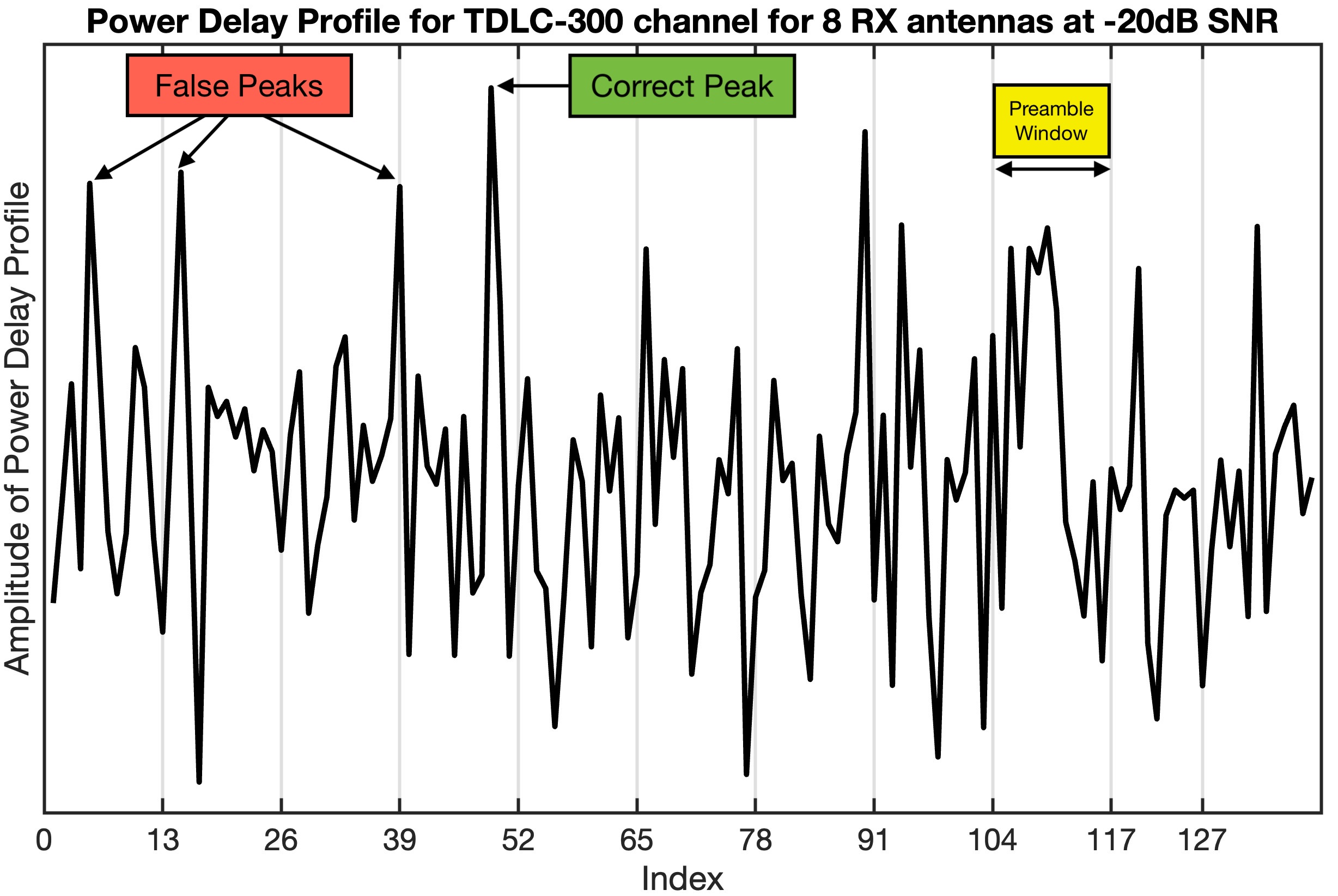}
    \caption{Power Delay Profile (PDP) of the correlation at the receiver, showing that it can be difficult to set a threshold due to multiple false peaks at low SNR.}
    \label{fig: Power delay profile}
\end{figure}

An example of the PDP of the correlation is shown in Fig.~\ref{fig: Power delay profile} for the case of EGC across 8 receiver antennas. We observe that at low SNRs, even after combining values across 8 RX antennas, there are still some false peaks. Setting up a threshold is very difficult in such scenarios. If the threshold is set too high, then there will be a missed detection. If the threshold is too low, there will be a lot of false peaks.


\subsubsection{AI/ML approaches}

Several attempts at PRACH detection (user detection, timing advance estimation and false peak detection) using AI/ML have been made. A two-stage AI/ML framework is described in~\cite{quek} where a first classifier predicts the number of UEs likely to collide. Separate classifiers are designed for TA estimation. Other AI/ML models such as k-Nearest Neighbors (kNN)~\cite{modina2019machine, zehra2022proactive}, Naive Bayes~\cite{modina2019machine, zehra2022proactive}, and Decision Tree Classification (DTC)~\cite{zehra2022proactive} have shown to improve the True/False classification performance. A further gain is achieved through ensemble learning methods~\cite{zehra2022proactive} and the models show 97\% user detection accuracy at $10$ dB and $20$ dB SNRs. 

An initial attempt at RAPID detection and TA estimation is made by~\cite{9951528}. However, this work does not consider impairments such as fading or noise, which are important for any practical receiver. Our previous work~\cite{singh2024enhancements} also developed AI/ML classifier models for RAPID detection and TA estimation. The receiver proposed in this paper is an improvement upon our previous work in terms of model complexity. 
The AI/ML classifier described in~\cite{choi2023structured} achieves good user detection performance using CDP as model input but it is limited to AWGN scenarios. In this paper, we show that in the presence of fading channels, CDP actually performs worse than PDP. 

The hybrid receiver proposed in this paper has an NN that can detect the presence or absence of users, without the need for thresholds. The NN works for all the possible base sequences. It also works in the case of multiple users, provided there is no collision of users within a preamble window. The NN can be trained offline, deployed in the BS and retrained periodically to adjust the weights to the changing scenarios. In an RA scenario, even if there are many users in a given cell, the probability of multiple users trying to access the network simultaneously is very low. So it makes sense to optimize the receiver for a single user instead of multiple users. Different from other works, we have evaluated the model with over-the-air data captured from the IIT Madras 5G testbed~\cite{5gtbiitm}. A further distinction of our work is that we have tried to explain why the model takes a specific decision. We have used SHapley Additive exPlanations (SHAP)~\cite{Shap} (which works on a game theoretic approach) to evaluate the model's decision-making.

\begin{table*}
    \centering
    \caption{Comparison of  various works (Conventional and Machine Learning based)}
    \begin{tabular}{|p{2cm}|p{3cm}|p{3cm}|p{3cm}|p{4cm}|} \hline 
         \textbf{Paper} &  \textbf{AI/ML or Non AI/ML} &  \textbf{Assumptions} & \textbf{Probability of Detection} & \textbf{Proposed hybrid receiver}\\ \hline 
         ~\cite{zehra2022proactive}&  AI/ML& Real data from AZCOM Technology & 97\% at 10 dB and 20 dB SNR & 99.9\% at 10 dB \& 99.94\% at 20 dB SNR for TDLC-300 channel\\ \hline 
         ~\cite{choi2023structured}& AI/ML & Uses CDP for Preamble detection \& assumes pre-compensation of Timing Advance before giving to NN&  & We have shown that Power Delay Profile (PDP) based NN performs better than CDP \\ \hline 
         ~\cite{multi_threshold_journal}& Non AI/ML & Enhanced receiver with multi-thresholding technique & 97.60\% at -10 dB in AWGN channel, 2 RX antennas with the knowledge of TA. & 99.98 \% at -10 dB in AWGN channel, 2 RX antennas, without the knowledge of TA \\ \hline
         Conventional receiver (Correlation-based)& Non AI/ML & MATLAB inbuilt 3GPP compliant receiver~\cite{Matlab_5g} &  70.41\% at -10 dB SNR \& 94.31\% at -5 dB SNR for TDLC300 channel & 90.3\% at -10 dB SNR \& 98.76\% at -5 dB SNR for TDLC300 channel  \\ \hline
    \end{tabular}
    \label{tab:comparison_table}
\end{table*}

\section{Proposed hybrid PRACH receiver}
\begin{figure*}[h!]
    \centering
    \includegraphics[width=0.75\textwidth]{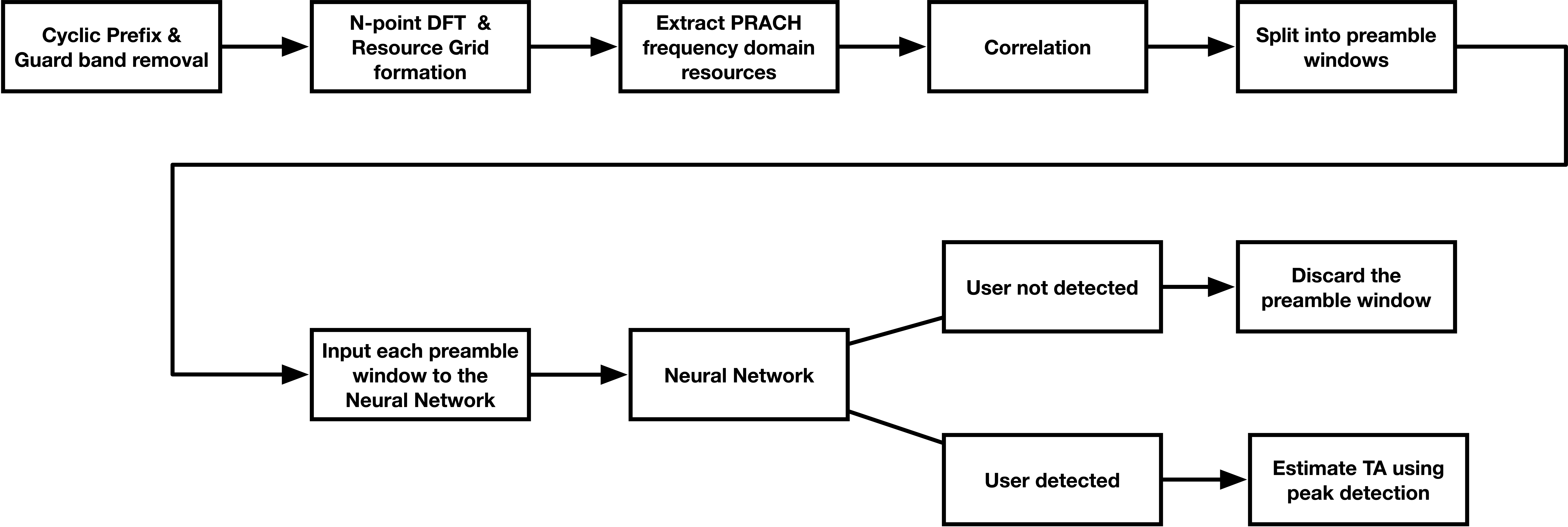}
    \caption{The proposed AI/ML based hybrid architecture for the PRACH receiver.}
    \label{fig: The proposed AI/ML based hybrid architecture for PRACH receiver}
\end{figure*}

Fig. \ref{fig: The proposed AI/ML based hybrid architecture for PRACH receiver} shows the proposed hybrid PRACH receiver. Initially, the data is extracted from the Resource Grid in the frequency domain. Subsequently, the extracted data is correlated with the base sequence and the PDP is calculated based on Eq.~\eqref{eq: pdp} The PDP is then separated into various preamble windows, and fed as separate input instances to the NN. The output prediction of the NN is the presence or absence of a user. If the NN detects the user in a preamble window, the RAPID corresponding that particular window can be identified. The window is then passed on to a peak detection module to find the timing advance. If the NN does not detect the user, this data in that window is discarded.

\subsection{Neural Network Architecture}
\begin{figure}[h!]
    \centering
    \includegraphics[width=0.5\textwidth]{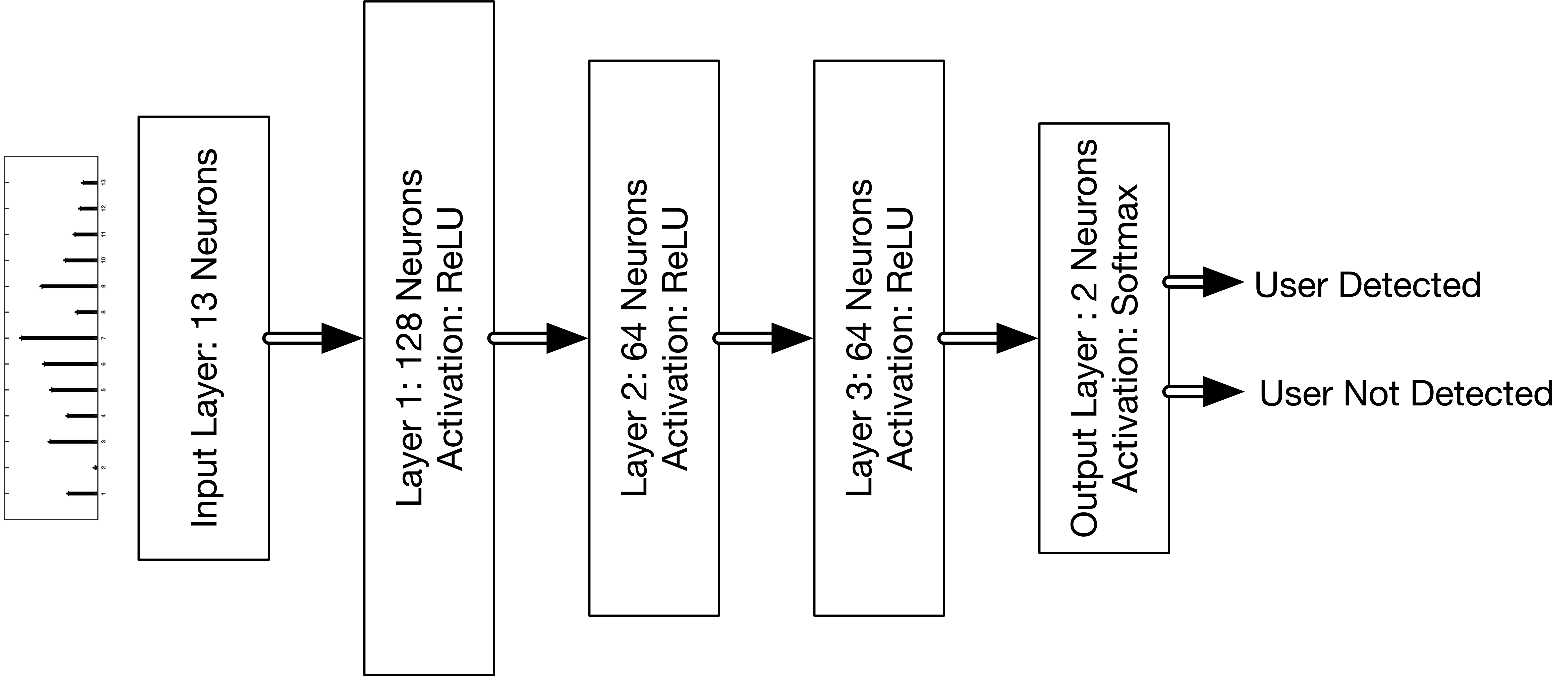}
    \caption{The proposed NN architecture for PRACH user detection.}
    \label{fig: The proposed NN architecture for PRACH user detection.}
\end{figure}

For the problem of user detection, we have used a simple Fully Connected NN. The architecture of the NN model is shown in Fig.~\ref{fig: The proposed NN architecture for PRACH user detection.}. The input to the model is the $N_{CS}$ length preamble window extracted from the PDP as shown in Eq.~\eqref{eq: pdp}. An example model input can be seen in Fig.~\ref{fig:Input_shap}. The model has three hidden layers consisting of 128, 64 and 64 neurons respectively. The output layer has two neurons, indicating whether the user is detected or not. 

\begin{figure}[h!]
    \centering
    \includegraphics[width=0.25\textwidth]{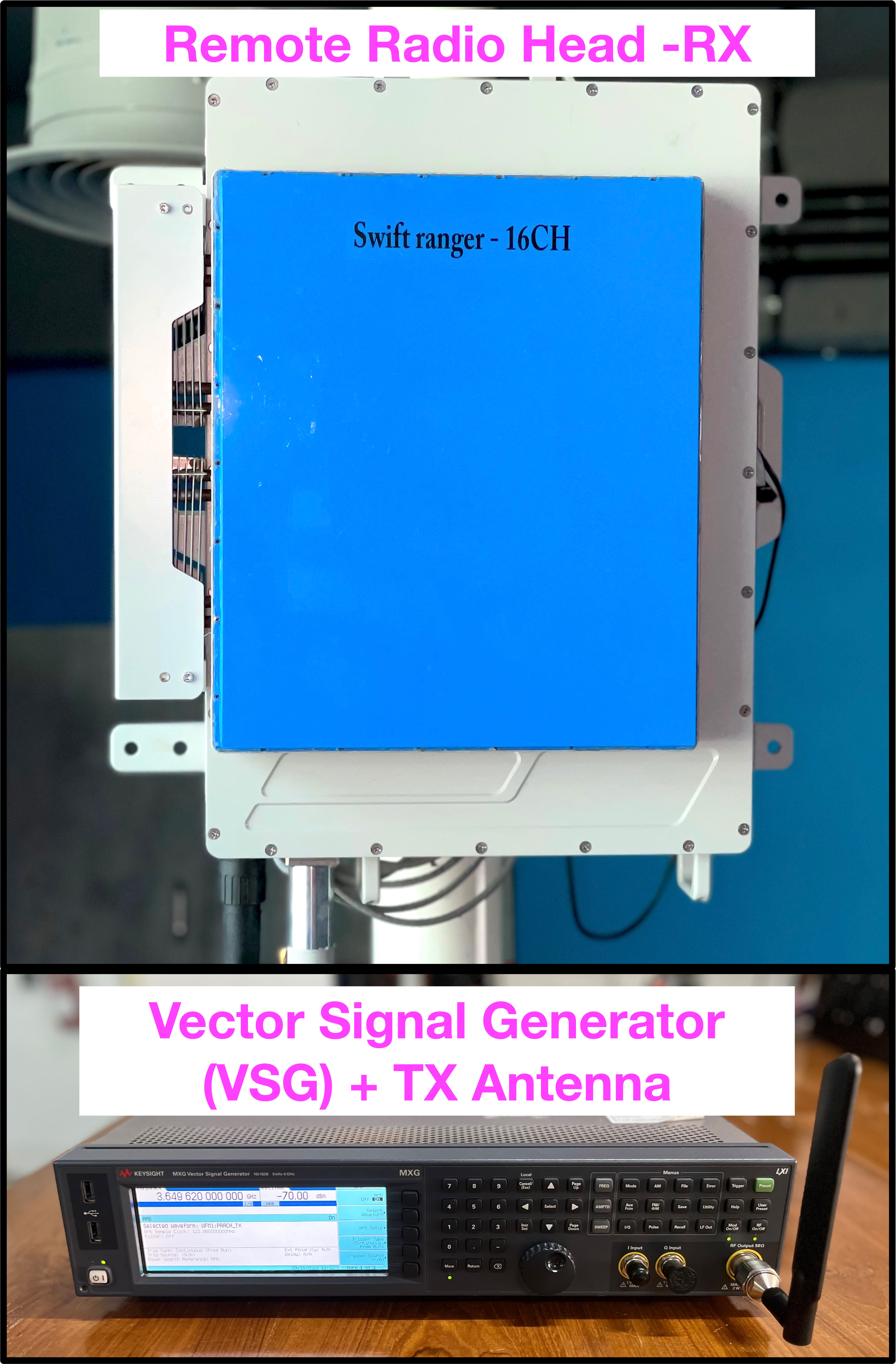}
    \caption{IIT-Madras 5G testbed setup with the Remote Radio Head used as a receiver and the VSG used as a transmitter.}
    \label{fig: hw_setup}
\end{figure}

\subsection{Dataset Generation}

\subsubsection{Simulated datasets}
For the training and testing of the NN model, we have generated datasets using MATLAB for the SNR range of $-20$ dB to $20$ dB in the steps of $5$ dB. The dataset contains 10000 instances for each SNR. During training, the data across all SNRs is combined. A train-test split of 75-25\% is used.

\subsubsection{Over-the-Air (OTA) Hardware captured datasets}
We have also tested the AI/ML model using real-world data from a base station at the IIT Madras 5G testbed to see how well the model works with actual hardware impairments such as those from DACs (Digital-to-Analog converters), ADCs (Analog-to-Digital converters), up/down converters, amplifiers, etc. Fig.~\ref{fig: hw_setup} shows the hardware setup used for capturing the real data. The transmitter and the receiver are set up one meter apart in a line-of-sight scenario. We used an N5182B Vector Signal Generator with a commercial omnidirectional wideband monopole antenna to transmit the signal over the air. The center frequency used is 3.64962 GHz, which lies in the N78 band for 5G NR. We have used an in-house 5G-compliant base station to receive the signal. One receiver antenna is used to receive the data. On the receiver side, the data undergoes various hardware impairments arising from multiple components. The data is quantized and passed through the different components in the receiver chain. Frequency domain data is collected from the receiver and then pre-processed in MATLAB. After correlation with the base sequence the data is divided into various preamble windows. This data is used for AI/ML model training and testing.

\section{Results}
This section shows the performance analysis of the proposed hybrid receiver for PRACH in terms of the user detection accuracy and the Timing Advance estimation accuracy.

\begin{figure}[!ht]
        \centering
        \includegraphics[width=\linewidth]{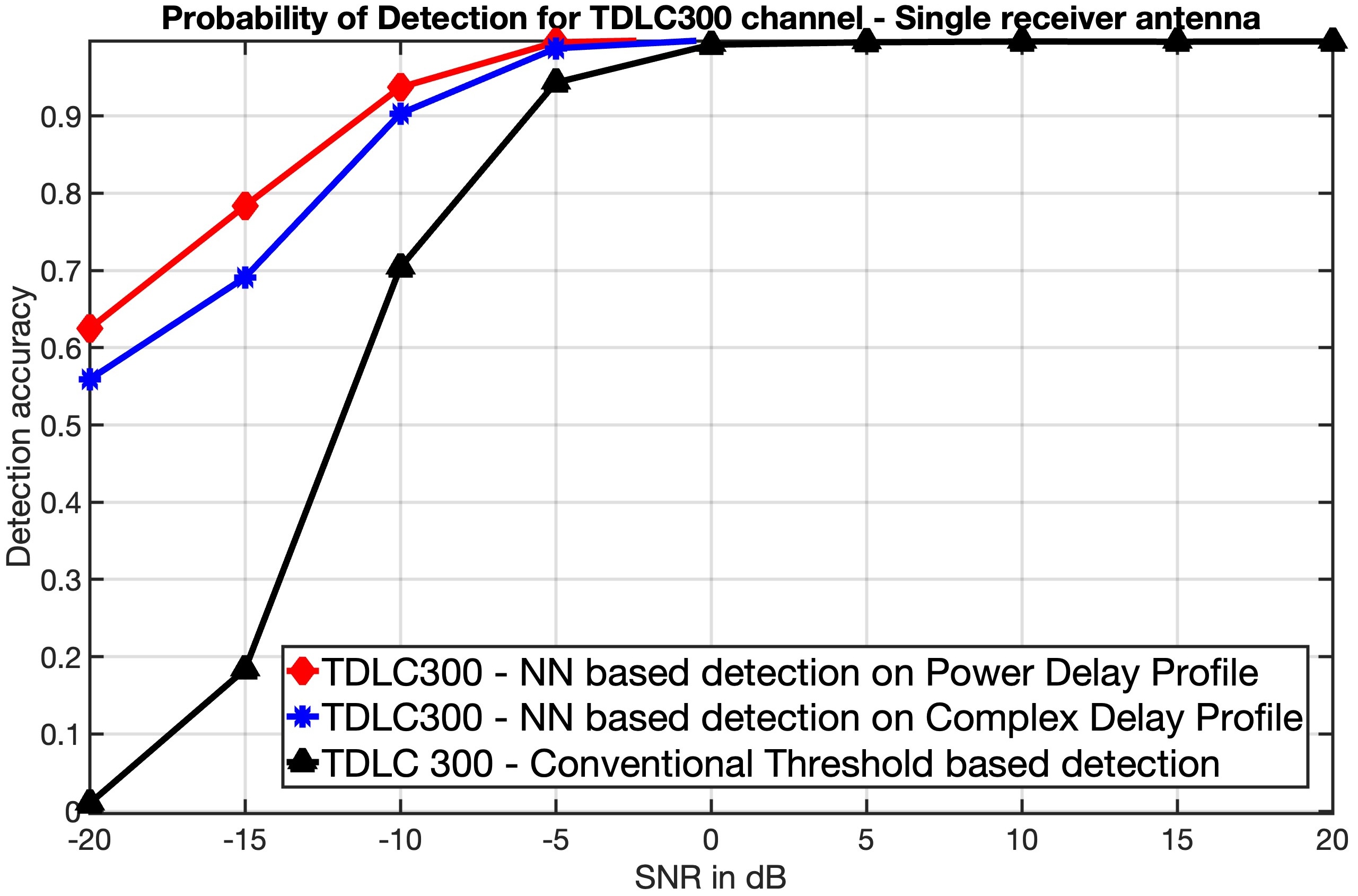}
        \caption{Comparison of user detection probability of AI/ML based receiver (PDP and CDP inputs) with conventional threshold based detection}
        \label{fig:Probability of User Detection for TDLC300 channel - Single receiver antenna}
    \end{figure}

In Fig.~\ref{fig:Probability of User Detection for TDLC300 channel - Single receiver antenna}, we observe that the performance of the hybrid receiver with AI/ML is significantly better than the conventional correlation-based receiver. In the low SNR range, the conventional correlation-based receiver suffer from false detection and missed detections. These false peaks and missed detections depend on the predefined threshold at the receiver. This threshold is difficult to change in a deployed system in real-time. Within AI/ML, PDP based model input is better than CDP based model input. This is because the information about the user's presence is embedded within the correlation peak and not in the phase of the delay profile. Another problem with the CDP-based method is that it suffers from destructive interference when combining across multiple antennas. 

\begin{figure}[!ht]
        \centering
        \includegraphics[width=\linewidth]{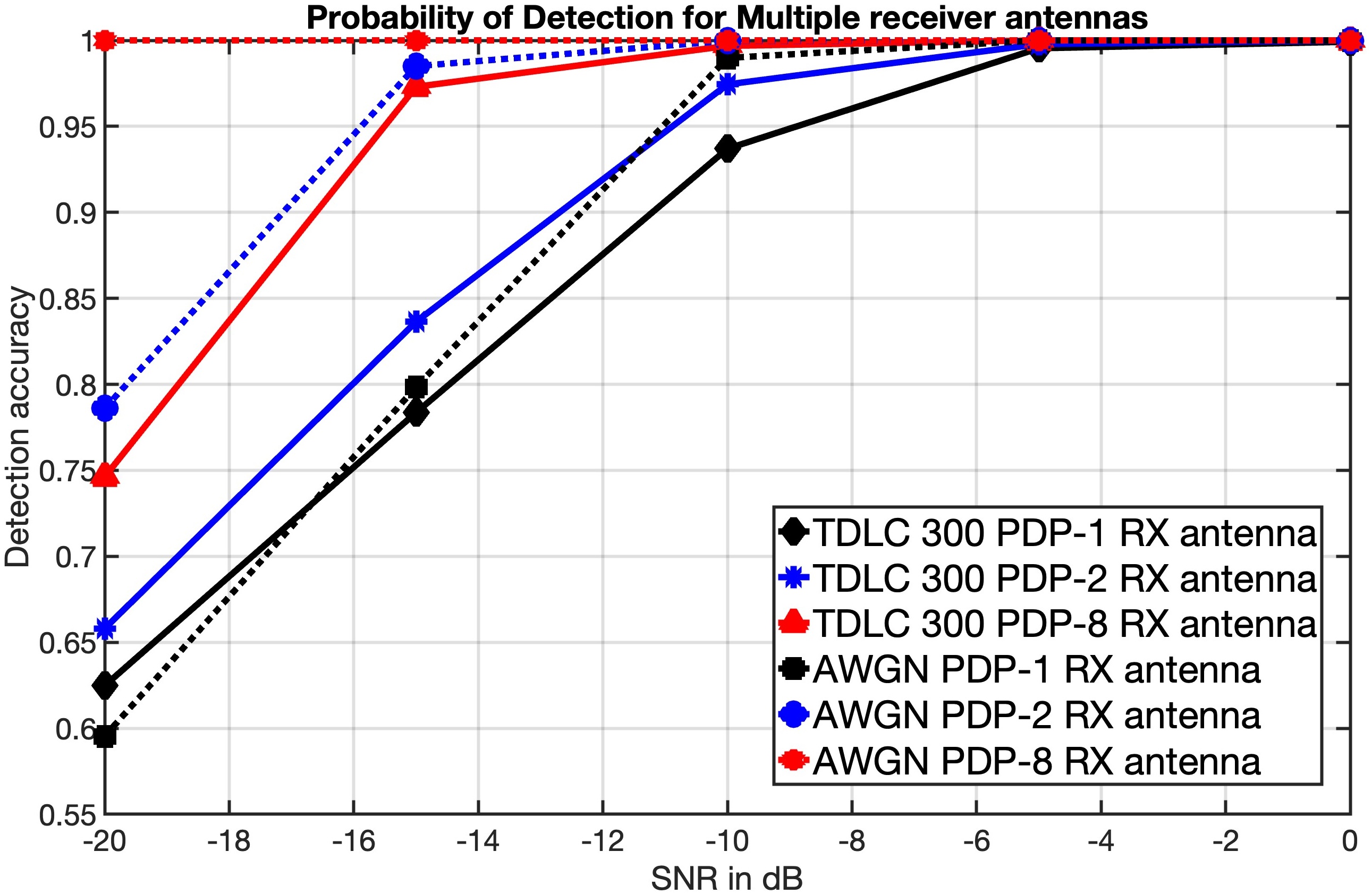}
        \caption{Comparison of user detection probability of the AI/ML based receiver (PDP inputs) across multiple receiver antennas.}
        \label{fig:MIMO PDP AWGN and TDLC 300}
    \end{figure}

Fig.~\ref{fig:MIMO PDP AWGN and TDLC 300} shows the performance of the PDP-based NN for multiple receiver antenna scenarios, with EGC, for AWGN and the TDLC300 channel. Here, the number of transmit antennas at the UE is 1, and the number of receiver (BS) antennas considered is 1, 2, and 8. We can see that detection accuracy in AWGN is better than in the case of TDLC300 fading. Furthermore, especially for the lower SNR cases, the performance with multiple antennas is better than that with a single antenna.



\begin{figure}[!ht]
        \centering
        \includegraphics[width=\linewidth]{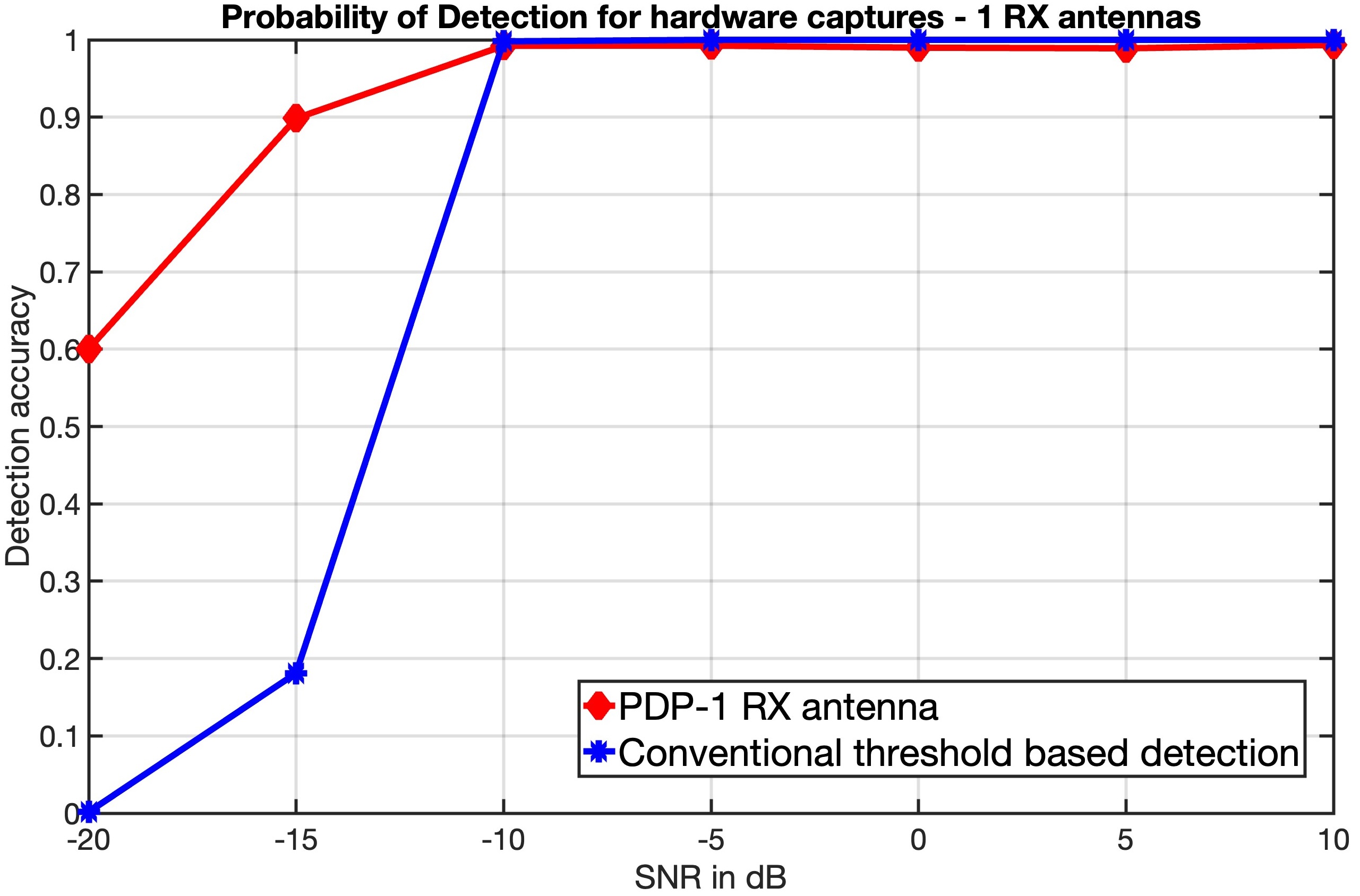}
        \caption{Performance of the proposed AI/ML model for hardware captures taken at 5G Testbed IIT Madras.}
        \label{fig:HW_performance}
    \end{figure}

Fig.~\ref{fig:HW_performance} shows the performance comparison of the NN and the correlation-based receiver for the hardware captures (Note that the model is retrained using the hardware dataset). Particularly in the low SNR range, the NN performs better because the correlation-based receiver has a static threshold due to which, there are a lot of missed detections and false peaks. On the other hand, the  NN looks at the peak and the neighboring values in the decision-making process, making it more robust in low SNR and multipath environments. 

After the user is detected in a preamble window, a peak detection module is used to find the Timing Advance (TA). The estimated timing advance is the shift of the peak from the rightmost point in the preamble correlation window. The rotation in the correlation is induced by two things: one is the delay caused by the distance from the UE to the BS, and another is the multipath. In this experiment, we used a TDLC300 channel, which introduced a random amount of delay in the signal, on top of the delay induced by the distance between the UE and the BS. Since it is difficult to characterize the delay introduced by the TDLC300 channel, it is difficult to obtain the ground truth labels for the TA. Here, we have defined two different ground truth labels for comparison purposes. There are as follows:
\begin{enumerate}
    \item delay due to distance + average delay due to channel
    \item delay due to distance + average delay due to channel + 1 TA error tolerance (because we don't know the exact delay and the 1 TA error can be easily corrected by the cyclic prefix of the subsequent PUSCH transmission, even if the estimated TA value is wrong by one value).
\end{enumerate}

\begin{figure}[!ht]
        \centering
        \includegraphics[width=\linewidth]{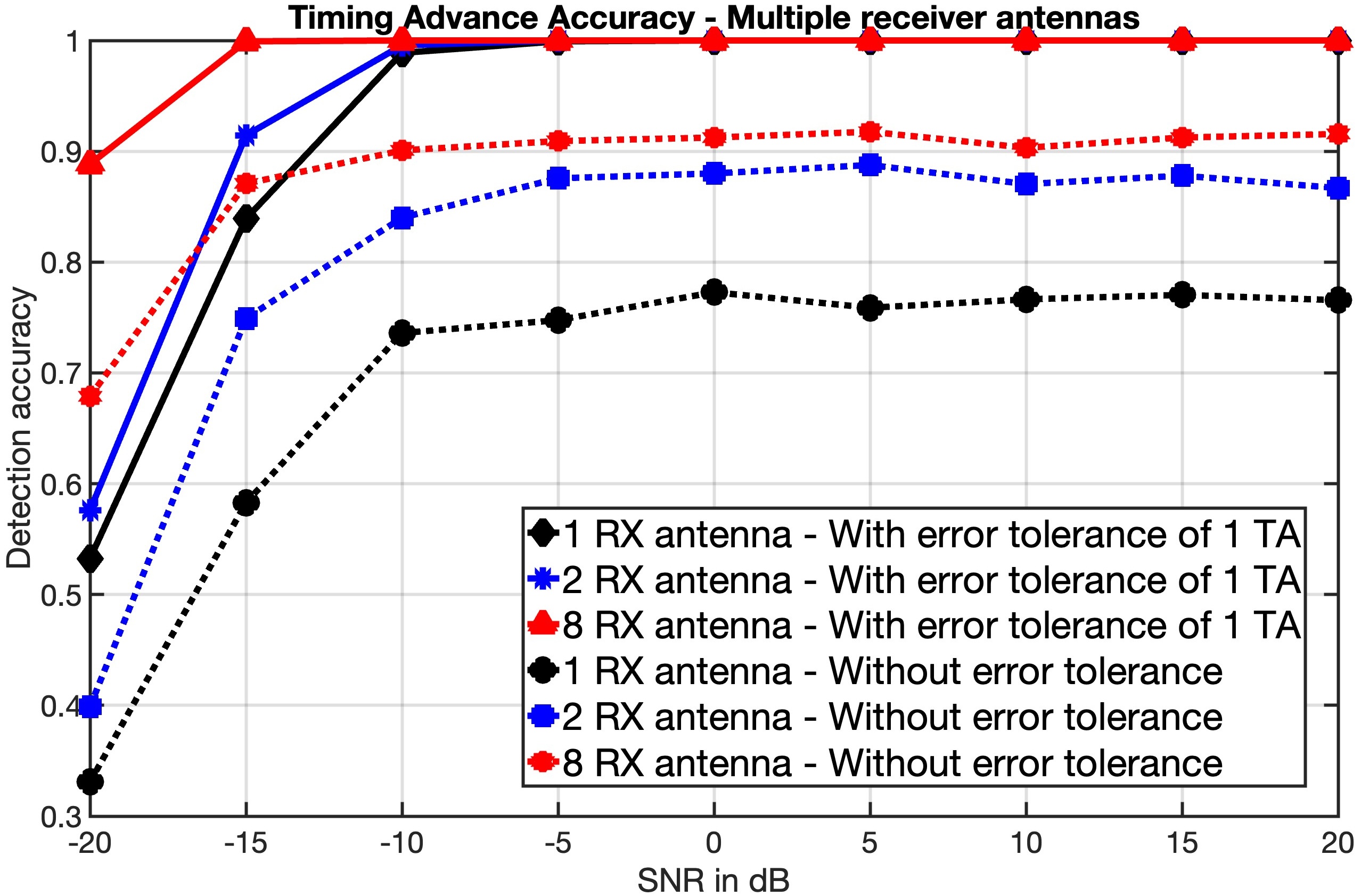}
        \caption{Peak detection based Timing Advance accuracy for different ground truth label and different number of receiver antennas.}
        \label{fig:Peak detection based Timing Advance accuracy for MIMO.}
    \end{figure}

We show the accuracy of these two ground truth labels for various numbers of RX antennas. We see that the performance is better when we allow the error of one TA as shown in Fig.~\ref{fig:Peak detection based Timing Advance accuracy for MIMO.}. This case is more practical since we don't know the exact delay introduced by the channel.
    
\subsection{Explainability of the model output}
\begin{figure}[!ht]
    \centering
    \begin{subfigure}[b]{0.5\textwidth}
        \centering
        \includegraphics[width=\textwidth]{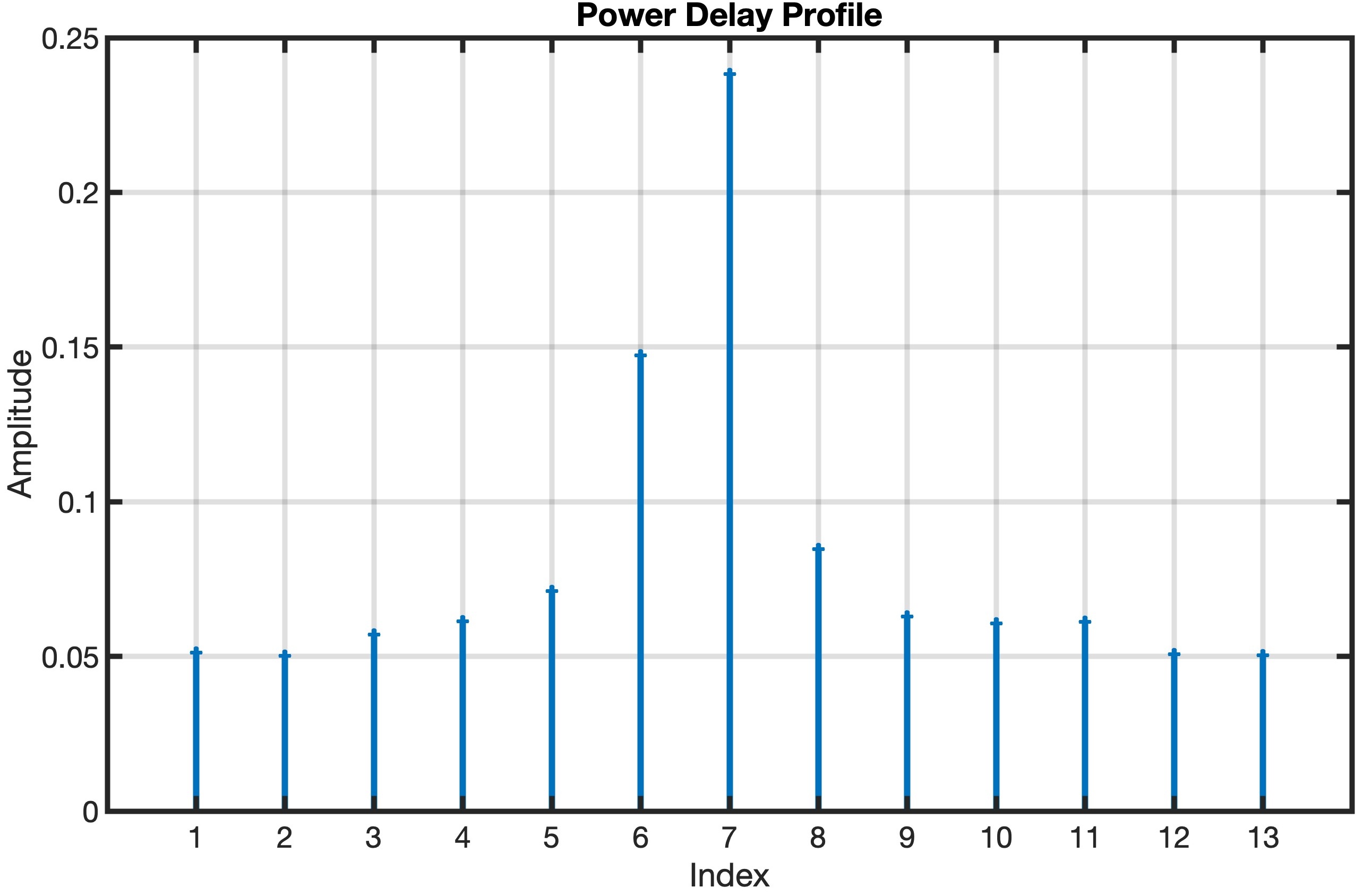} 
        \caption{}
        \label{fig:Input_shap}
    \end{subfigure}
    \hfill
    \begin{subfigure}[b]{0.5\textwidth}
        \centering
        \includegraphics[width=\textwidth]{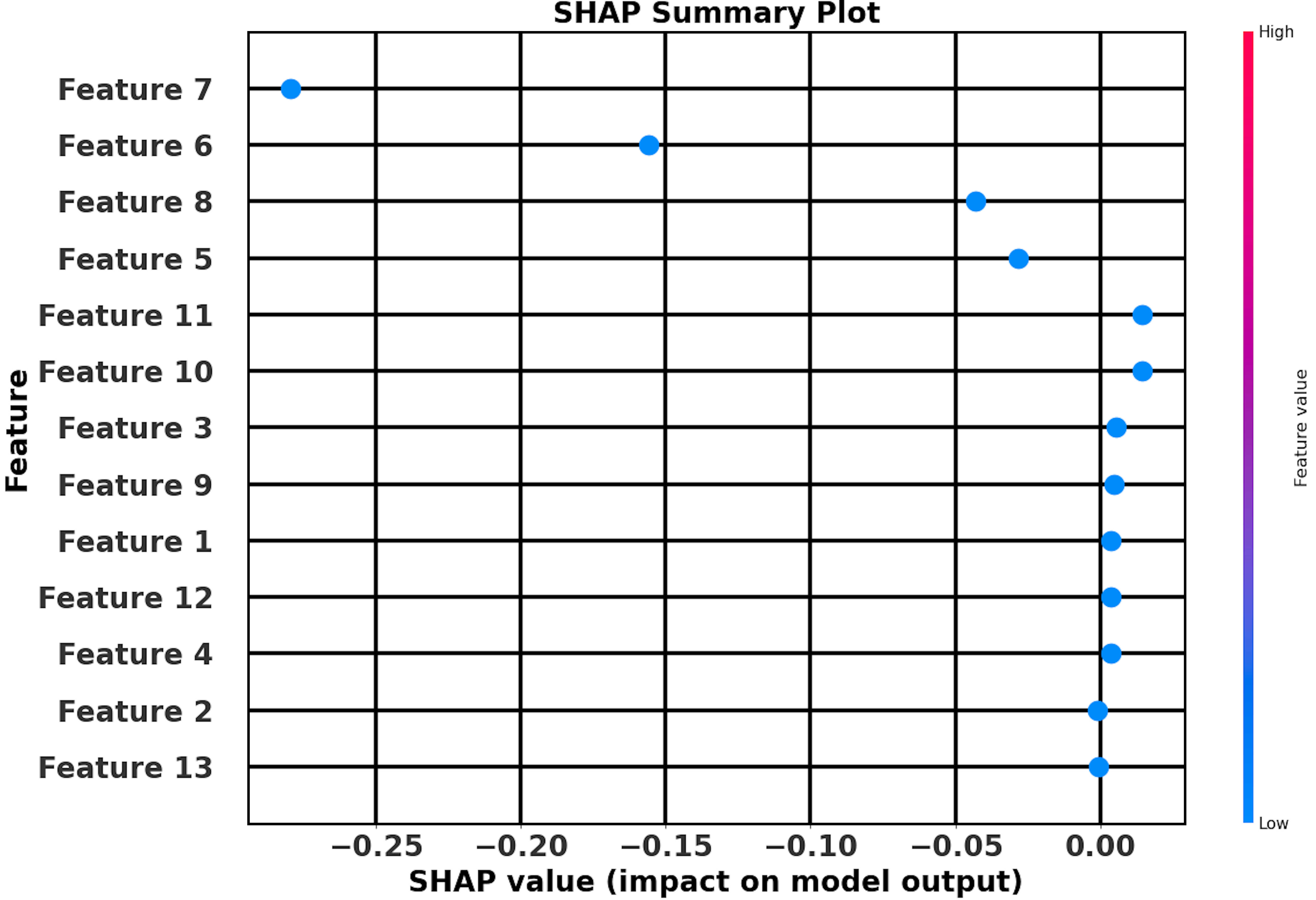} 
        \caption{}
        \label{fig:Output_shap}
    \end{subfigure}
    \caption{(a) Example AI/ML model input showing the PDP with the peak at the $7^{th}$ location (b) SHAP explanation summary showing that the model is using not only the peak, but also the values around the peak caused due to multipath.}
    \label{fig:main}
\end{figure}

In addition to detection accuracy and TA estimation, we have looked into the explainability of the AI/ML model using SHAP. It is a game theoretic approach that explains the output of any trained machine learning model. Using the SHAP library in Python, we understand which of the inputs had more impact on the model output (presence or absence of a user, in this case). In other words, we want to know which parts of the input helped the model predict the presence or absence of a user. Naturally, the index where the peak is present should carry more weight compared to the other indices in the PDP (input). In the case of multipath channels, the peak is scattered due to various channel paths. In this case, the neighboring indices also carry significant weight. Fig.~\ref{fig:Input_shap} shows an example input to the model, and Fig.~\ref{fig:Output_shap} shows the corresponding SHAP values. In the example shown in Fig.~\ref{fig:Input_shap}, the highest peak is at the $7^{th}$ location, followed by the $6^{th}$, $8^{th}$ location and so on. This pattern is also reflected in the SHAP values in Fig.~\ref{fig:Output_shap}.

\subsection{Model Complexity}
The neural network proposed in this paper has around $28000$ Floating Point Operations Per Second (FLOPS). The correlation operation has around $5800$ FLOPS. For all $64$ preamble windows across all $7$ base sequences, the total FLOPS are of the order of $1.8$ Million FLOPS. This is a significant improvement over our previous work in which two separate NN models (For RAPID and TA) have to run independently leading to around $593$ Million FLOPS. There are two reasons for the lower complexity of the proposed receiver. The first is due to the use of correlation instead of the frequency domain data as model input. Another is the hybrid architecture, which removes the need for a separate NN for TA estimation. This reduced complexity makes the receiver more practical for deployment. Future work could include optimizations such as model pipelining that can significantly reduce inference time, especially when processing multiple iterations of the same model for each preamble window.



    

\section{Conclusion}
This paper presents a hybrid receiver consisting of an AI/ML model for user detection and conventional peak detection for timing advance estimation. This model can easily integrate with current implementations of correlation based receivers in 5G base stations and does not require setting up thresholds. Simulation results under TDLC MIMO channel models as well as hardware captures from a real 5G testbed base station, show the proposed receiver's performance to be better than the conventional methods. Using explainability frameworks such as SHAP, we provide some insights into the inner workings of the AI/ML model. Future work could extend to actual hardware deployment of the proposed receiver and also to very low power devices such as Ambient IoT. 

\section*{Acknowledgment}
The authors would like to thank the Department of Telecommunications (DOT), India,  for funding the 5G Testbed project and the Ministry of Electronics and Information Technology (MeitY) for funding this work through the project "Next Generation Wireless Research and Standardization on 5G and Beyond".


\bibliographystyle{IEEEtran}
\bibliography{bibfile}

\begin{thebibliography}{10}
\providecommand{\url}[1]{#1}
\csname url@samestyle\endcsname
\providecommand{\newblock}{\relax}
\providecommand{\bibinfo}[2]{#2}
\providecommand{\BIBentrySTDinterwordspacing}{\spaceskip=0pt\relax}
\providecommand{\BIBentryALTinterwordstretchfactor}{4}
\providecommand{\BIBentryALTinterwordspacing}{\spaceskip=\fontdimen2\font plus
\BIBentryALTinterwordstretchfactor\fontdimen3\font minus \fontdimen4\font\relax}
\providecommand{\BIBforeignlanguage}[2]{{%
\expandafter\ifx\csname l@#1\endcsname\relax
\typeout{** WARNING: IEEEtran.bst: No hyphenation pattern has been}%
\typeout{** loaded for the language `#1'. Using the pattern for}%
\typeout{** the default language instead.}%
\else
\language=\csname l@#1\endcsname
\fi
#2}}
\providecommand{\BIBdecl}{\relax}
\BIBdecl

\bibitem{singh2024enhancements}
R.~Singh, A.~K. Yerrapragada, J.~S. Keshav, and R.~K. Ganti, ``Enhancements for 5{G} {NR} {P}rach {R}eception: {A}n {AI/ML} {A}pproach,'' in \emph{2024 Wireless Telecommunications Symposium (WTS)}.\hskip 1em plus 0.5em minus 0.4em\relax IEEE, 2024, pp. 1--6.

\bibitem{3gpp_38_211}
3GPP, ``{Physical Channels and Modulation},'' {3rd Generation Partnership Project (3GPP)}, Technical Specification (TS) 38.211, 2018, version 15.3.0.

\bibitem{pham2019proposed}
T.~A. Pham and B.~T. Le, ``{A proposed preamble detection algorithm for 5G-PRACH},'' in \emph{2019 International Conference on Advanced Technologies for Communications (ATC)}.\hskip 1em plus 0.5em minus 0.4em\relax IEEE, 2019, pp. 210--214.

\bibitem{kamata2021detection}
K.~Kamata, M.~Sawahashi, and Y.~Kishiyama, ``{Detection Probability of PRACH Preamble for NR in 3GPP TDL Channel Models},'' in \emph{2021 IEEE VTS 17th Asia Pacific Wireless Communications Symposium (APWCS)}.\hskip 1em plus 0.5em minus 0.4em\relax IEEE, 2021, pp. 1--5.

\bibitem{hu2012method}
Y.~Hu, J.~Han, S.~Tang, H.~Gao, Y.~Su, and J.~Shi, ``{A method of PRACH detection threshold setting in LTE TDD femtocell system},'' in \emph{7th International Conference on Communications and Networking in China}.\hskip 1em plus 0.5em minus 0.4em\relax IEEE, 2012, pp. 408--413.

\bibitem{threshold_2}
S.~Kim, K.~Joo, and Y.~Lim, ``{A delay-robust random access preamble detection algorithm for LTE system},'' in \emph{2012 IEEE Radio and Wireless Symposium}, 2012, pp. 75--78.

\bibitem{multi_threshold_journal}
Y.~T. Song and S.~W. Choi, ``On the {S}tructured {D}esign {M}ethodology of {E}ffective {PRACH} {D}etection: {M}ulti-{S}tage {T}hresholding {P}erspective,'' \emph{IEEE Access}, vol.~12, pp. 32\,192--32\,199, 2024.

\bibitem{quek}
H.~S. Jang, H.~Lee, T.~Q.~S. Quek, and H.~Shin, ``{Deep Learning-Based Cellular Random Access Framework},'' \emph{IEEE Transactions on Wireless Communications}, vol.~20, no.~11, pp. 7503--7518, 2021.

\bibitem{modina2019machine}
N.~Modina, R.~Ferrari, and M.~Magarini, ``{A machine learning-based design of PRACH receiver in 5G},'' \emph{Procedia Computer Science}, vol. 151, pp. 1100--1107, 2019.

\bibitem{zehra2022proactive}
S.~S. Zehra, M.~Magarini, R.~Qureshi, S.~M.~N. Mustafa, and F.~Farooq, ``Proactive approach for preamble detection in 5g-nr prach using supervised machine learning and ensemble model,'' \emph{Scientific reports}, vol.~12, no.~1, p. 8378, 2022.

\bibitem{9951528}
R.~Fang, H.~Chen, and W.~Liu, ``{Deep Learning-Based PRACH Detection Algorithm Design and Simulation},'' in \emph{2022 2nd International Conference on Frontiers of Electronics, Information and Computation Technologies (ICFEICT)}, 2022, pp. 505--511.

\bibitem{choi2023structured}
S.~W. Choi \emph{et~al.}, ``On the {S}tructured {D}esign for {E}fficient {M}achine {L}earning {B}ased {PRACH} {P}reamble {D}etection,'' in \emph{2023 14th International Conference on Information and Communication Technology Convergence (ICTC)}.\hskip 1em plus 0.5em minus 0.4em\relax IEEE, 2023, pp. 1017--1021.

\bibitem{5gtbiitm}
``{IIT Madras 5G Testbed},'' \url{http://www.5gtbiitm.in/}, {Accessed: 2024-15-10}.

\bibitem{Shap}
\BIBentryALTinterwordspacing
S.~M. Lundberg and S.~Lee, ``{A} {U}nified {A}pproach to {I}nterpreting {M}odel {P}redictions,'' \emph{CoRR}, vol. abs/1705.07874, 2017. [Online]. Available: \url{http://arxiv.org/abs/1705.07874}
\BIBentrySTDinterwordspacing

\bibitem{Matlab_5g}
``{Matlab 5G Toolbox},'' \url{https://in.mathworks.com/products/5g.html}, {Accessed: 2024-15-10}.

\end{thebibliography}
\end{document}